\documentclass[%
 reprint,
 amsmath,amssymb,
 aps, showpacs, showkeys
]{revtex4-2}

\usepackage{graphicx}
\usepackage{dcolumn}
\usepackage{bm}

\begin{document}


\title{Quantum entanglement in mixed-spin trimer: effects of a magnetic field and heterogeneous $g$-factors}

\author{Zhirayr Adamyan}
\affiliation{Laboratory of Theoretical Physics,
          Yerevan State University,
         1 Alex Manoogian Str., 0025 Yerevan, Armenia}
 \affiliation{CANDLE, Synchrotron Research Institute, 31 Acharyan Str., 0040 Yerevan, Armenia}

\author{Vadim Ohanyan}
 \affiliation{Laboratory of Theoretical Physics,
          Yerevan State University,
         1 Alex Manoogian Str., 0025 Yerevan, Armenia}
 \affiliation{CANDLE, Synchrotron Research Institute, 31 Acharyan Str., 0040 Yerevan, Armenia}

\date{\today}

\begin{abstract}
Mixed spin-(1/2,1/2,1) trimer with two different Land\'{e} $g$-factors and two different exchange couplings is considered. The main feature of the model is non-conserving magnetization. The Hamiltonian of the system is diagonalized analytically. We presented a detailed analysis of the ground state properties, revealing several possible ground state phase diagrams and magnetization profiles. The main focus is on how non-conserving magnetization affects quantum entanglement. We have found that non-conserving magnetization can bring to the continuous dependence of the entanglement quantifying parameter (negativity) on magnetic field within the same eigenstate, while for the case of uniform $g$-factors it is a constant. The main result is an essential enhancement of the entanglement in case of uniform couplings for one pair of spins caused by an arbitrary small difference in the values of $g$-factors. This enhancement is robust and brings to almost 7-fold increasing of the negativity. We have also found weakening of entanglement for other cases. Thus, non-conserving magnetization offers a broad opportunity to manipulate the entanglement by means of magnetic field.
\end{abstract}

\maketitle


\section{\label{sec:level1}Introduction}

Quantum entanglement, due to its critical role in quantum communication and information processing, has recently attracted more and more attention \cite{ent,guhne,bephys54,bephys76,ekert}. It is an essential ingredient of quantum teleportation \cite{bephys70, bouwlondon, chophys95, jinnature, baurphys, bjphysrev}, quantum computing \cite{kokphysmod, kimphysrev, brodphysrev}, and quantum cryptography \cite{vedral89, avellaphysrev}.  Moreover,  comprehension of quantum entanglement has facilitated a progress in understanding black holes using quantum field theory methods. Entanglement also provides a new perspective for studying quantum phase transitions and collective phenomena in many-body systems and condensed matter physics.
Recently, studying entanglement features of quantum spin clusters or molecular magnets \cite{molmag, kahn} has became very popular, as there are firm evidences that molecular magnets can be promising materials for realizing qubits for quantum information technologies \cite{los01, ste08, ses15}. In numerous papers, which have been published recently, various aspects of quantum entanglement in few-body spin systems have been figured out \cite{ana11, abg11, ana12, str14, car18, sou19, sou20, ada20, cen20, eki20, kar20, str20, gal21, gal21b, gal22, ben22, zad22, zhe22}.

There are different methods of quantifying quantum entanglement using special entanglement measures \cite{horodecki}. The entanglement measure, exploited in the present paper, is the so-called "negativity" \cite{vidal}. The principal advantage of negativity is its universality, i. e. possibility of its formulation and calculation in terms of the reduced two-particle density matrix independent of the physical nature of the system. Moreover, it can be efficiently exploited not only for the zero temperature entanglement (pure quantum entanglement) but, for the finite $T$ as well (thermal entanglement).

The field of quantum molecular magnetism is rapidly advancing, with researchers pursuing both fundamental and practical applications of magnetic molecules and materials. This interdisciplinary field draws on physics, chemistry and nanotechnology, with specific applications in molecular spintronics, quantum technologies, metal-organic frameworks, and molecular spin qubits \cite{molmag, kahn, los01, ste08, ses15}. From formal point of view, molecular magnets can be considered as models of a few interacting spins (magnetic moments) with a variety of peculiar magneto-thermal properties. Such materials are usually referred to as zero-dimensional magnets.  Some magnetic materials with structure of molecular crystal may also contain a one-dimensional arrays of interacting spins in the form of coordination polymers. Normally, magnetization processes of zero-dimensional magnets at low enough temperature reflect their finite spectrum in a straightforward way. A typical zero-temperature magnetization curve of the models of few interacting spins is a system of plateaus with jumps between them. Each plateau corresponds to the certain eigenstate with fixed value of magnetization, which is the ground state at given values of magnetic field. However, the magnetization profile can be changed crucially if the model possesses a non-conserving magnetization operator \cite{sou19, ada20, cen20, str05, vis09, van10, bel14, oha15, tor16, tor18, var19, kro20, pan20}. In this case magnetization is not a good quantum number, thus, its value for a given eigenstate can be non-constant. This, in its turn, leads to a continuous dependance of the magnetization expectation value on the magnetic filed within the same eigenstate \cite{oha15}. For the spin clusters or molecular magnets with non-conserving magnetization, in virtue of possible continuous dependence of magnetization within a single eigenstate, low-temperature magnetization curves profiles have a lot in common with magnetization processes in real many-body models. It is worth mentioning that for the systems with non-conserved $z$-component of total  spin, $S_{tot}^z$, magnetization is non-conserving as well. However, even if $\left[\mathcal{H}, S_{tot}^z\right]=0$ the magnetization can not commute with the system Hamiltonian, provided different spins have different $g$-factors \cite{bel14, oha15, var19, kro20}. Even for the simplest isotropic model with the Hamiltonian
\begin{eqnarray}\label{ham_gen}
\mathcal{H}=\sum_{i, j} J_{ij}\mathbf{S}_i\mathbf{S}_j-B\mathcal{M}^z, \;\; \mathcal{M}^z=\sum_i g_i S_i^z
\end{eqnarray}
one has
\begin{eqnarray}\label{com}
\left[\mathcal{H}, \mathcal{M}^z\right]=i\sum_{i,j}J_{ij}(g_i-g_j)\left(S_i^xS_j^y-S_i^yS_i^x\right).
\end{eqnarray}
Thus, even if a single pair of spins in the lattice has $g_i\neq g_j$, then the magnetization operator does not commute with the Hamiltonian.

In this work, we consider a mixed spin trimer with two spin-1/2 and one spin-1 magnetic ions, with two different exchange couplings and two different but isotropic Land\'{e} $g$-factors. We consider the case when ion with spin 1 and one of the ions with spin 1/2 have $g$-factor equal to $g_1$ while the second spin-1/2 ion has $g_2$ (See Fig. \ref{fig1}). Also, the distribution of exchange couplings is a bit unusual, as we assume that the one of the ions with spin-1/2 interacts with spin-1 and another spin-1/2 with the same exchange constant, $J_1$. The main reason for such configuration of the system is its highest possible symmetry, facilitating the analytical finding of the Hamiltonian eigenvalues and eigenstates. The paper deals with the problem of finding the optimal parameters to create entangled states and control them by external magnetic field. The main goal is to figure out how non-conserving magnetization affects entanglement measure like negativity. For all quantities of interest analytical consideration is possible. Our main finding is that inhomogeneous $g$-factors for appropriate values of other parameters lead to enhancement of entanglement with respect to the case when all spins are taken to have the same $g$-factors.

The paper is organized as follows, in the Second section we formulate the model and present its exact spectrum and eigenstates. The next Third section is devoted to zero-temperature phase diagrams and magneto-thermal properties of the system. In section IV we calculate quantum ($T=0$) negativity and present the plots of its magnetic field behaviour. The paper ends with Conclusion.

\section{\label{sec:level2}The Model and its exact solution}
\begin{figure}[h]
\resizebox{0.5\textwidth}{!}{
 \includegraphics{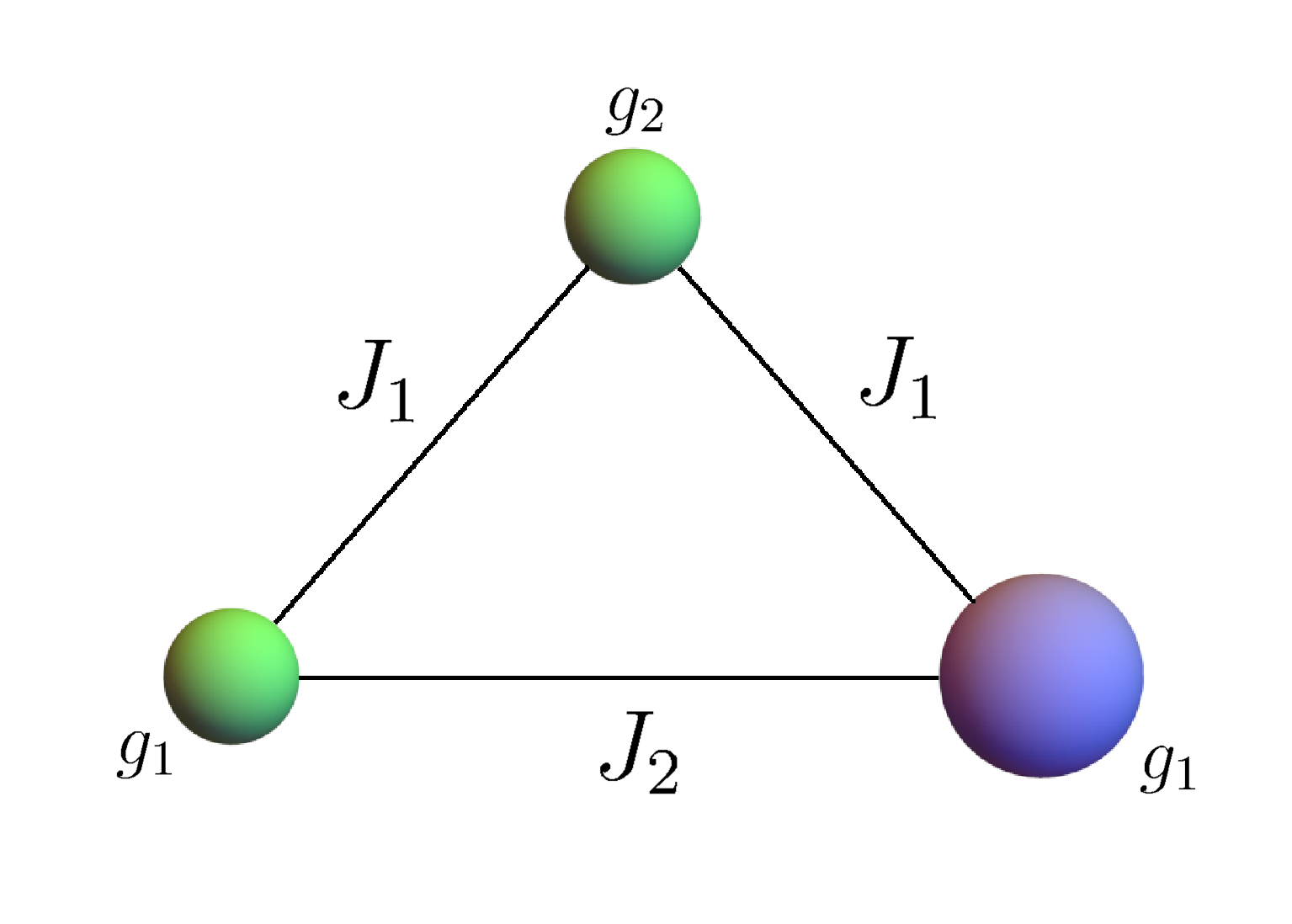} 
}
\caption
{Symbolic picture of triangular trinuclear mixed-spin magnetic molecule with spins 1/2, 1/2 (small balls), and 1 (large ball). The Lang\`{e} $g$-factors for one spin-1/2 ion and spin-1 ion are supposed to be the same ($g_1$), whereas the $g$-factor for the second spin-1/2 is different from them, $g_2$.}
\label{fig1}
\end{figure}

Let us consider a model of trimetallic heterogeneous molecular magnet with spins $1/2, 1/2$ and $1$ in triangular geometry. We also suppose the exchange coupling between two spin-$1/2$ ions, $J_2$, is a different from the other two, $J_1$. We focuss on the effects of non-conserving magnetization, caused by the non-uniformity of $g$-factors. Therefore, the model under consideration includes two magnetic ions with $g$-factors equal to $g_1$ (one of the spin-1/2 ions and spin-1 ion) and another spin-1/2 ion with $g$-factor equal to $g_2$ (See Fig. \ref{fig1}).

The Hamiltonian of the model has the following form:
\begin{equation}\label{ham}
\mathcal{H} = J_{1}\left(\boldsymbol s_{1}\boldsymbol s_{2}+\boldsymbol s_{2}\boldsymbol S_{3}\right)+J_{2}\boldsymbol s_{1}\boldsymbol S_{3}-B\left(g_{1}s_{1}^z+g_{2}s_{2}^z+g_{1}S_{3}^z\right)
\end{equation}
where $\boldsymbol s_{a}$, $a=1, 2$, are spin-1/2 operators, and  $\boldsymbol S_{3}$ stands for spin-1 operators. Exchange interaction between all pairs of spins is supposed to be isotropic. Due to presence of magnetic ions with two different $g$-factors, the magnetization operator does not commute with the Hamiltonian,
\begin{eqnarray}
&&[\mathcal{H},M^{z}] = iJ_{1}g_{-}\left(s_{1}^x s_{2}^y - s_{1}^y s_{2}^x+ s_{2}^yS_{3}^x -  s_{2}^xS_{3}^y \right), \nonumber \\
&&M^z=g_{1}s_{1}^z+g_{2}s_{2}^z+g_{1}S_{3}^z,\nonumber \\
&& g_-=g_1-g_2.
\end{eqnarray}
However, $S_{tot}^z=s_1^z+s_2^z+S_3^z$ conserves.
The Hamiltonian (\ref{ham}) can be diagonalized analytically. The twelve eigenvalues are
\begin{eqnarray}\label{eigval}
{E}_{1,2}&=&\frac{1}{4}\left(3J_{1}+2J_{2}\mp2B(3g_{1}+g_{2})\right),\nonumber \\
{E}_{3,4}&=&\frac{1}{4}\left(J_{1}-4J_{2}\mp2B(g_{1}+g_{2})\right), \\
{E}_{5,6}&=&\frac{1}{4}\left(-J_{1}-4J_{2}\mp2\sqrt{B^{2}g_{-}^2+J_1^2 }\right),\nonumber \\
{E}_{7,8}&=&\frac{1}{4}\left(-J_{1}+2J_{2}\mp2\sqrt{B^{2}g_{-}^2+4J_1^2}\right),\nonumber \\
{E}_{9,10}&=&\frac{1}{4}\left(-J_{1}+2J_{2}-4Bg_{1}\mp2Q^+\right),\nonumber \\
{E}_{11,12}&=&\frac{1}{4}\left(-J_{1}+2J_{2}+4Bg_{1}\mp2Q^-\right), \nonumber \\
\mbox{where} \; \; \; Q^\pm &=& \sqrt{(Bg_{-}\pm J_{1})^2+3J_{1}^2}. \nonumber
\end{eqnarray}
Non-linear dependence of some eigenvalues on the magnetic field is the direct consequence of non-commutativity of the magnetization operator and the Hamiltonian. The corresponding eigenvectors respectively are
\begin{widetext}
  \begin{eqnarray}\label{eigvec}
&& |\psi_{1,2}\rangle= \left|\pm\frac{1}{2},\pm\frac{1}{2},\pm1\right\rangle, \\
 && |\psi_{ 3,4}\rangle=\frac{1}{\sqrt{3}}
\left(\pm\sqrt{2}\left|\mp\frac{1}{2},\pm\frac{1}{2},\pm1\right\rangle \mp\left|\pm\frac{1}{2},\pm\frac{1}{2},0\right\rangle\right),  \nonumber \\
&&  |\psi_{ 5,6}\rangle= \frac{1}{\sqrt{3(1+M^{2}_{\pm})}}
\left(M^{\pm}\left(\sqrt{2}\left|\frac{1}{2},\frac{1}{2},-1\right\rangle- \left|-\frac{1}{2},\frac{1}{2},0\right\rangle\right)+\sqrt{2}\left|-\frac{1}{2},-\frac{1}{2},1\right\rangle- \left|\frac{1}{2},-\frac{1}{2},0\right\rangle\right), \nonumber \\
&&|\psi_{ 7,8}\rangle= \frac{1}{\sqrt{3(4+K^{2}_{\mp})}}
\left(K^{\mp}\left(\left|\frac{1}{2},\frac{1}{2},-1\right\rangle+\sqrt{2}\left|-\frac{1}{2},\frac{1}{2},0\right\rangle\right)+
2\left|-\frac{1}{2},-\frac{1}{2},1\right\rangle+ 2\sqrt{2}\left|\frac{1}{2},-\frac{1}{2},0\right\rangle\right), \nonumber \\
  &&|\psi_{ 9,10}\rangle= \frac{1}{\sqrt{3+G^{2}_{\pm}}}
\left(\sqrt{2}\left|\frac{1}{2},\frac{1}{2},0\right\rangle-
G^{\pm}\left|\frac{1}{2},-\frac{1}{2},1\right\rangle+ \left|-\frac{1}{2},\frac{1}{2},1\right\rangle\right), \nonumber \\
&&|\psi_{ 11,12}\rangle= \frac{1}{\sqrt{3+U^{2}_{\mp}}}
\left(\left|\frac{1}{2},-\frac{1}{2},-1\right\rangle+
U^{\mp}\left|-\frac{1}{2},\frac{1}{2},-1\right\rangle+ \sqrt{2}\left|-\frac{1}{2},-\frac{1}{2},0\right\rangle\right),\nonumber
\end{eqnarray}
\end{widetext}
where the following notations are adopted:
\begin{eqnarray}
&&M^\pm=\frac{-Bg_{-}\pm\sqrt{B^2g_{-}^2+J_{1}^2}}{J_{1}}, \\
&&K^\pm=\frac{Bg_{-}\pm\sqrt{B^2g_{-}^2+4J_{1}^2}}{2J_{1}}, \nonumber \\
&&G^\pm=\frac{Bg_{-}+J_{1}\pm Q^+}{J_{1}}, \nonumber \\
&&U^\pm=\frac{Bg_{-}-J_{1}\pm Q^-}{J_{1}}, \nonumber
\end{eqnarray}
and the basis vectors obey the standard notations, $\left|s_1^z, s_2^z, S_3^z\right\rangle$.
As has been mentioned above, they are eigenvectors of $S_{tot}^z$ operator as well, having eigenvalues form $-2$ to $2$. Non-conserving magnetization operator leads to situation where only four eigenvectors of the system, $\left|\psi_1\right\rangle-\left|\psi_4\right\rangle$, have fixed value of magnetic moment, in contrast to the others. However, the expectation value of the magnetization operator can be evaluated for each of them:
\begin{eqnarray}\label{mags}
&&M_{1,2}=\pm\frac{3g_{1}+g_{2}}{2}, \;M_{3,4}=\pm\frac{g_{1}+g_{2}}{2}, \\
&&M_{5,6}=-\frac{(M_{\pm}^2-1)g_{-}}{2(M_{\pm}^2+1)}, \; M_{7,8}=-\frac{(K_{\mp}^2-4)g_{-}}{2(K_{\mp}^2+4)}, \nonumber \\
&&M_{9,10}=\frac{3g_{+}+G_{\pm}^2(3g_{1}-g_{2})}{2(G_{\pm}^2+3)}, \nonumber \\
&&M_{11,12}=-\frac{3g_{+}+2U_{\mp}^2(3g_{1}-g_{2})}{2(2U_{\mp}^2+3)}, \nonumber \\
&&g_+=g_1+g_2. \nonumber
\end{eqnarray}
Due to non-commutativity of the magnetization operator and the Hamiltonian of the system majority of the eigenstates have magnetic field and $J_1$ dependent expectation values of the magnetization  \cite{bel14, oha15}. Interestingly, dependence on $J_2$ does not occur. The way we introduced non-equal exchange interaction in the model is a special one. The part of the Hamiltonian which describes interaction between first and third spins turns to be conserved,
\begin{eqnarray}\label{J2_comm}
[\mathcal{H},J_{2}\boldsymbol s_{1}\boldsymbol S_{3}] = 0.
 \end{eqnarray}
Thus,  the spectrum of a system with non-zero $J_2$ and the spectrum of the system with $J_2=0$ differ from each other by the constant shift proportional to $J_2$. Therefore, magneto-thermal properties of the system in triangular shape ($J_{2} \ne 0 $) and in linear geometry ($J_{2} = 0 $) are to a great extent identical.
Interestingly, not all eigenvectors from Eq.(\ref{eigvec}) are continuous under the limit restoring conservation of magnetization, $g_2\rightarrow g_1$. This is a quite common situation in the spin clusters with non-uniform $g$-factors \cite{oha15}. In case of our model six eigenvectors, corresponding to $S^z=0, \pm 1$ do not change continuously under the limit of uniform $g$-factors. For the triangular mixed spin cluster with $g_1=g_2$ the counterparts of eigenvectors $|\psi_{3,4}\rangle$, $|\psi_6\rangle$, $|\psi_7\rangle$ $|\psi_9\rangle$ and $|\psi_{11}\rangle$ have the following form:
\begin{widetext}
\begin{eqnarray}\label{psi_0}
&&|\psi_{3,4}\rangle_0=\frac 12 \left(\left|\frac 12, -\frac 12, \pm 1\right\rangle+\left|-\frac 12, \frac 12, \pm 1\right\rangle-\sqrt 2\left|\pm\frac 12, \pm\frac 12, 0\right\rangle\right),\\
&& |\psi_ 6\rangle=\frac{1}{\sqrt{2}}\left( \left|-\frac{1}{2},-\frac{1}{2},1\right\rangle-\left|\frac{1}{2},\frac{1}{2},-1\right\rangle \right), \nonumber \\
&&  |\psi_7\rangle=\frac{1}{\sqrt{2}}\left(\left|-\frac{1}{2},\frac{1}{2},0\right\rangle-\left|\frac{1}{2},-\frac{1}{2},0\right\rangle \right), \nonumber \\
&&|\psi_9\rangle_0=\frac{1}{\sqrt{2}} \left(\left|-\frac 12, \frac 12, 1\right\rangle-\left|\frac 12, -\frac 12, 1\right\rangle\right), \nonumber\\
&&|\psi_{11}\rangle_0=\frac{1}{\sqrt{2}} \left(\left|-\frac 12, \frac 12, -1\right\rangle-\left|\frac 12, -\frac 12, -1\right\rangle\right). \nonumber
\end{eqnarray}
\end{widetext}
The rest of the eigenvectors change continuously under $g_2\rightarrow g_1$.

 Last but not least, unusual choice of the interaction constants distribution have to be explained. This is quite an extraordinary suggestion that the exchange interaction constant between two spin-1/2 ions are the same as between spin-1 and spin-1/2 ions. It is worth mentioning, that in principle this can be the case, as least within certain numerical range. Thus, speaking about possible experimental realization of the model, one should have in mind numerical proximity of two exchange constants. From the practical point of view, we aimed to have the model which admits analytical solution in a possible simple way. Analysing all possible combinations of $g$-factors and coupling we arrive at the model given by the Hamiltonian (\ref{ham}), which due to conservation of $S_{tot}^z$, $J_{2}\boldsymbol s_{1}\boldsymbol S_{3}$ and discrete symmetry $\boldsymbol s_{1}\leftrightarrow\boldsymbol S_{3}$ have block-diagonal form with four one-dimensional and four two-dimensional blocks. This is the most simple Hamiltonian among all possible $(1/2, 1/2, 1)$ spin mixed clusters with two $g$-factors.

\section{\label{sec:3}Ground states phase diagrams and magnetization processes}

\begin{figure}
  \resizebox{0.5\textwidth}{!}{
 \includegraphics{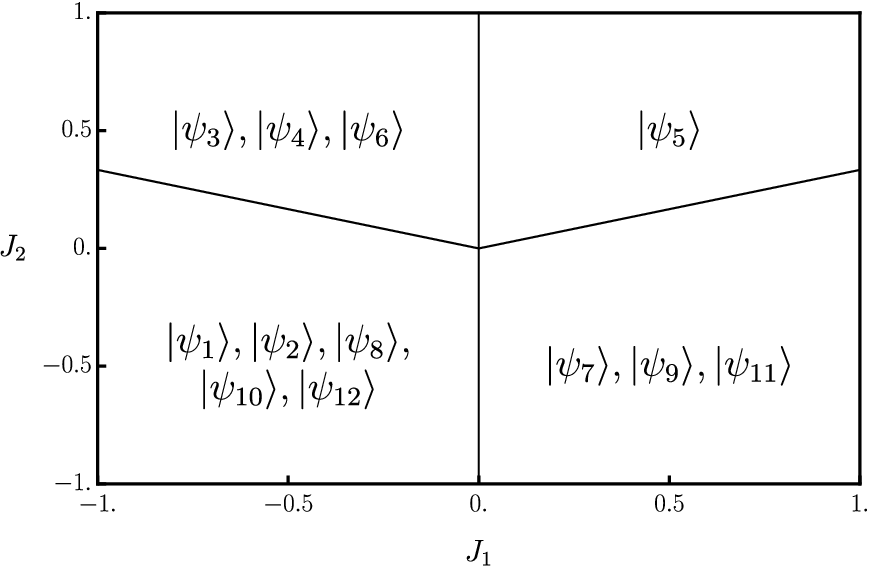} 
}
\vspace{-0.75cm}
\caption{Zero-field ground-states phase diagram.}
\label{fig_2}
\end{figure}

\begin{figure}
  \resizebox{0.5\textwidth}{!}{
 \includegraphics{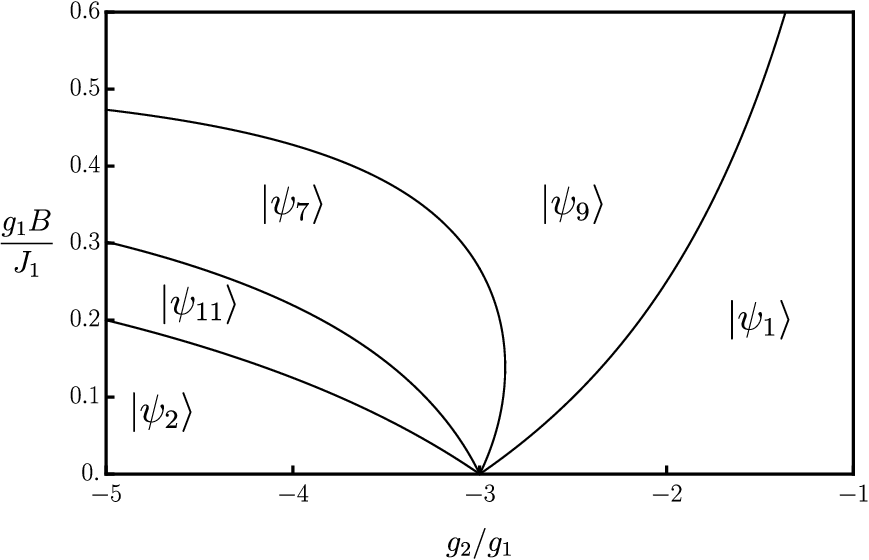} 
}
\vspace{-0.75cm}
\caption{Ground states phase diagram for $J_1=-1$ and arbitrary $J_2 \leq 0$.}
\label{fig_3}
\end{figure}

As the model under consideration has many parameters and possesses non-conserving magnetization operator, the variety of possible ground state phase diagrams is quite large.
First of all, one should distinguish the case of ferromagnetic, $J_1<0, J_2<0$, antiferromagnetic, $J_1>0, J_2>0$ and mixed $J_1J_2<0$ couplings. For the case of zero magnetic field the ground states phase diagram exhibits large degeneracy, except for the region where $|\psi_5\rangle$ is a ground state (See Fig. \ref{fig_2}). Obviously, the ground states and their properties are highly affected by the $g$-factors difference, $g_-$. Therefore, in this section we present the ground-state phase diagrams in $(g_2/g_1, Bg_1/J_1)$ or $(g_2/g_1, Bg_1/J_2)$ planes, depending on which coupling constant we set as an energy unit.
Without loss of generality one can consider five different types of ground state phase diagrams, given by the following conditions for the coupling constants: purely ferromagnetic, $J_1=-1$, $J_2\leq0$ (Fig. \ref{fig_3}); purely antiferromagnetic with equal couplings, $J_1=J_2=1$ (Fig. \ref{fig_5}); purely antiferromagnetic with non-equal couplings, $J_2>0, J_1>0$ (Fig. \ref{fig_7}); two cases of mixed coupling, $J_1=1$, $J_2 \leq 0$ (Fig. \ref{fig_9}) and $J_1<0$, $J_2=1$ (Fig. \ref{fig_11}).

\begin{figure}
   \centering
  \resizebox{0.5\textwidth}{!}{
 \includegraphics{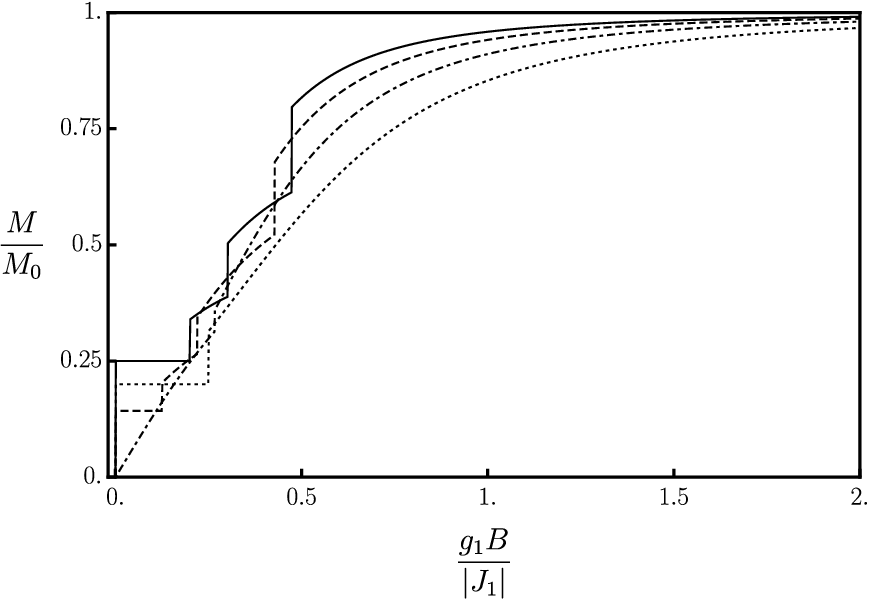} 
}
\vspace{-0.75cm}
\caption{ Magnetization processed for $J_1=-1$ and arbitrary $J_2\leq 0$ for $T/|J_1|=0.00001$ and $g_1=1$ for different negative values of $g_2$: $g_2=-5$ (solid); $g_2=-4$ (dashed) and $g_2=-3$ (dot-dashed). The value of the saturated magnetization, $M_0$, is $\frac{3g_1+|g_2|}{2}$.}
\label{fig_4}
\end{figure}

\subsubsection{Purely ferromagnetic couplings}
For the system under consideration three regions of ferromagnetic parameters can be specified: $J_{1} = J_{2}  < 0 $, $J_{1} < 0, J_{2} < 0, J_{1} \ne J_{2} $ and $ J_{1} < 0, J_{2} = 0  $. The former one corresponds to the case of linear spin cluster, which can be described as an limiting case of our model. Interestingly, due to the commutativity of the $J_2$-term with the Hamiltonian (Eq. (\ref{J2_comm})),  the ground state energies for all those cases differ from each other only by the same constant value $ 2J_{2}$, therefore all ground state phase diagram for ferromagnetic coupling have the same structure, presented in the Fig. \ref{fig_3}. To have non-trivial eigenstates in addition to the fully polarized one, $|\psi_1\rangle$, we have to investigate the region of negative $g$-factors. Speaking about negative $g$-factors one should bear in mind that this situation can be relevant rather for pseudospin \cite{chib}, mainly for the magnetic ions with large values of orbital moment and strong spin-orbit coupling. However, magnetic materials with negative or mixed $g$-factors are not very exotic \cite{tor18, kro20, ata08, chib12, yin13, yin24} and can be found among molecular magnets \cite{ata08} as well as among the low-dimensional quantum magnets \cite{yin13, yin24}. Moreover, theoretical investigations of the quantum and classical spin models with mixed $g$-factors reveal many unusual effects, like various partially ordered phases with disordered sublattices \cite{tor18,yin24}, or additional features in the spin structure factors \cite{kro20}.
  The very distinguishing feature of the phase diagram in Fig. \ref{fig_3} is the zero-field highly degeneracy. Moreover, there is no continuous transition between $B\neq 0$ ground states and zero-field ground-states manifold. Actually, at $B=0$ the following eigenstates become degenerate: $|\psi_1\rangle$, $|\psi_2\rangle$, $|\psi_8\rangle$, $|\psi_{10}\rangle$, $|\psi_{12}\rangle$. The same is valid for the special  point, $g_2=-3$.  Another interesting feature of the magnetization processes in this case which can be seen from the ground state phase diagram is asymptotic saturation. As one of the $S=1/2$ spins has negative $g$-factor in the case under consideration, the value of the total spin $S^z=s_1^z+s_2^z+S_3^z$ for the magnetically saturated state must be 1, whereas the saturated magnetization in the units of Bohr magnetons is equal to $\frac 12 \left(3g_1+|g_2|\right)$. In this case the eigenstates with $S^z=1$, $|\psi_9\rangle$, turns to be a saturated, or quasi-saturated ground state. Due to non-conserved magnetization the high-field ground state, $|\psi_9\rangle$, does not correspond to constant expectation value of the magnetization operator, on the contrary, as it can be seen form Eq. (\ref{mags}), the expectation value has explicit magnetic field dependence, leading to impossibility of reaching the saturation at finite values of magnetic field. This feature is also inherent to other two ground states, $|\psi_7\rangle$ and $|\psi_{11}\rangle$, appeared in the phase diagram from Fig. \ref{fig_3}. Thus, only two ground states $|\psi_1\rangle$ and $|\psi_2\rangle$ for purely ferromagnetic coupling have constant value of magnetization.
  Non-conserving magnetization, therefore, essentially modifies zero-temperature magnetization curves for finite spin clusters \cite{str05, bel14, oha15, tor18}. Instead of the typical step-like structure with plateaus corresponding to eigenstates with constant values of $M^z$ and jumps between them corresponding to transitions between eigenstates at given values of magnetic field, in some cases magnetization curves can contain continuous part of monotone magnetization change stemming out from the magnetic field dependant expectation values of $M^z$ for some eigenstates. To illustrate this behaviour we draw few plots of zero-temperature magnetization processes corresponding to phase diagram form Fig. \ref{fig_3} (See Fig. \ref{fig_4}).

\subsubsection{Purely antiferromagnetic couplings with $J_{1} = J_{2}$}
\begin{figure}
   \centering
  \resizebox{0.5\textwidth}{!}{
 \includegraphics{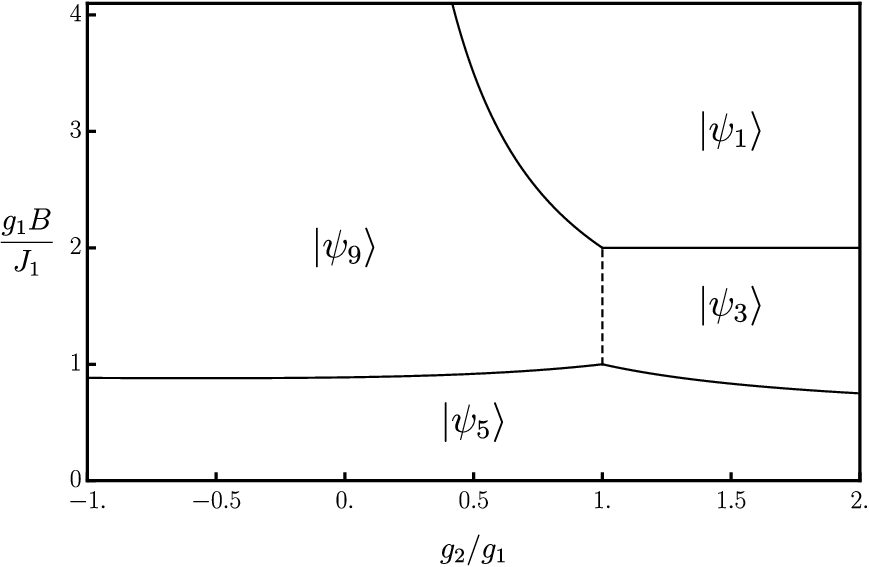} 
}
\vspace{-0.75cm}
\caption{ Ground states phase diagram for $J_1=J_2=1$.}
\label{fig_5}
\end{figure}

\begin{figure}[h!]
   \centering
  \resizebox{0.5\textwidth}{!}{
 \includegraphics{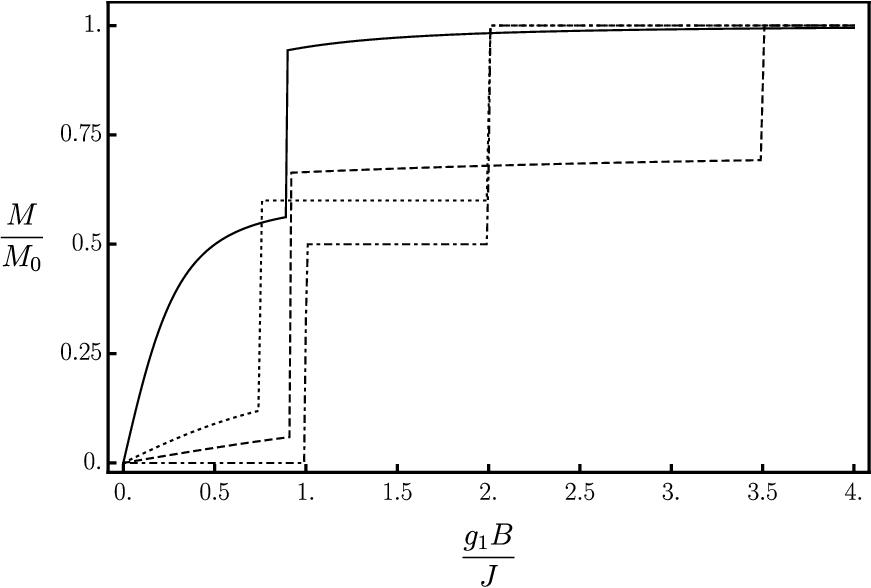} 
}
\vspace{-0.75cm}
\caption{ Magnetization processed for $J_1=J_2=1$ for $T/|J_1|=0.00001$ and $g_1=1$ for different values of $g_2$: $g_2=-2$ (solid); $g_2=1/2$ (dashed); $g_2=1$ (dot-dashed) and $g_2=2$ (dotted).}
\label{fig_6}
\end{figure}

For the case of uniform antiferromagnetic couplings, $J_1=J_2>0$, four ground states are possible, $|\psi_{1}\rangle, |\psi_{3}\rangle,|\psi_{5}\rangle$ and $|\psi_{9}\rangle$. In addition to that, when $g_2=g_1$ there is no continuous transition between the eigenvectors, corresponding to $|\psi_3\rangle$ and $|\psi_9\rangle$ eigenstates. Therefore, along dashed line in the ground states phase diagram in Fig. \ref{fig_5} the correct ground state in given by the degenerate doublet $|\psi_3\rangle_0, |\psi_9\rangle_0$ from the Eq. (\ref{psi_0}). Another interesting feature is zero-field ground states which turns to be $|\psi_5\rangle$ for all values of $g_2$ and is never degenerate in contrast to the previous case. The saturated state, the ground state for large enough values of magnetic field, is different for positive and negative values of $g_2$. For the positive values of the $g$-factors the ordinary saturated (maximally polarized) eigenstate, $|\psi_1\rangle$ is the saturated one, however, region of negative $g_2$ as in a previous case has $|\psi_9\rangle$ as high-field ground state. It is also worth mentioning, that only two ground states $|\psi_3\rangle$ and $|\psi_1\rangle$ have constant value of the magnetization. Additional intermediate ground state, $|\psi_3\rangle$ appears only when $g_2>g_1$. The zero-temperature magnetization curves illustrating the phase diagram are presented in Fig. \ref{fig_6}. Here one can see all the magnetization processes features we demonstrated in the ground state phase diagram, quasi-saturation for negative $g_2$ as well as regular saturation for $g_2 < 0$, intermediate states with constant and continuously growing magnetization and magnetization jumps.   \\\\

\begin{figure}
  \resizebox{0.5\textwidth}{!}{
 \includegraphics{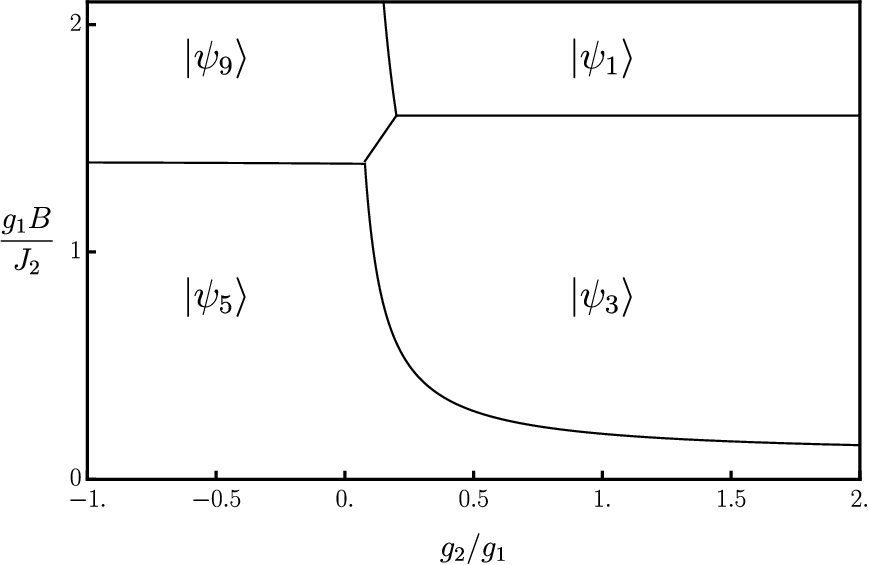} 
}
\vspace{-0.75cm}
\caption{ Ground-states phase diagram for $J_1=1/5$, $J_2=1$}
\label{fig_7}
\end{figure}

\begin{figure}[h!]
   \centering
  \resizebox{0.5\textwidth}{!}{
 \includegraphics{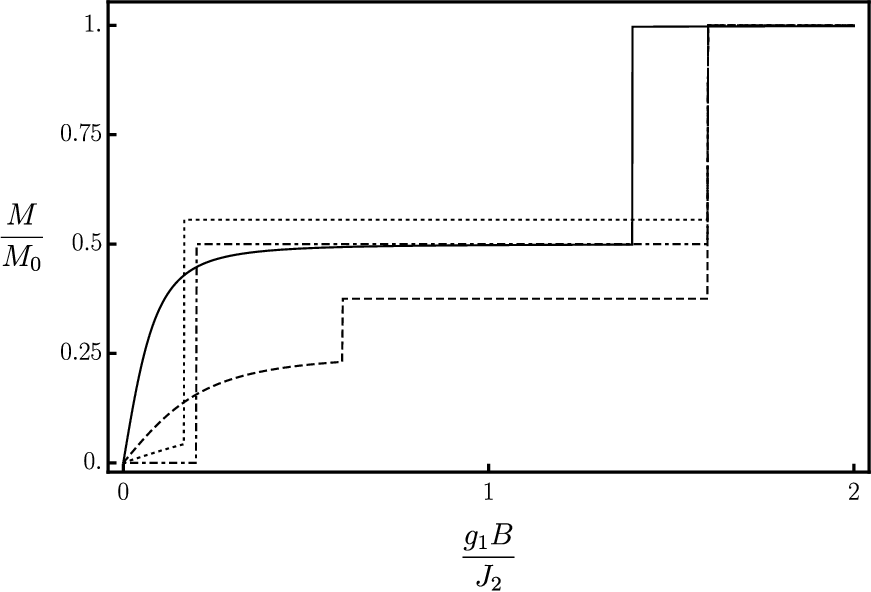} 
}
\vspace{-0.75cm}
\caption{ Magnetization processed for $J_1=1/51$, $J_2=1$, $T/|J_1|=0.00001$, $g_1=1$ for different values of $g_2$: $g_2=-1$ (solid); $g_2=1/5$ (dashed); $g_2=1$ (dot-dashed) and $g_2=3/2$ (dotted).}
\label{fig_8}
\end{figure}

\begin{figure}
   \centering
  \resizebox{0.5\textwidth}{!}{
 \includegraphics{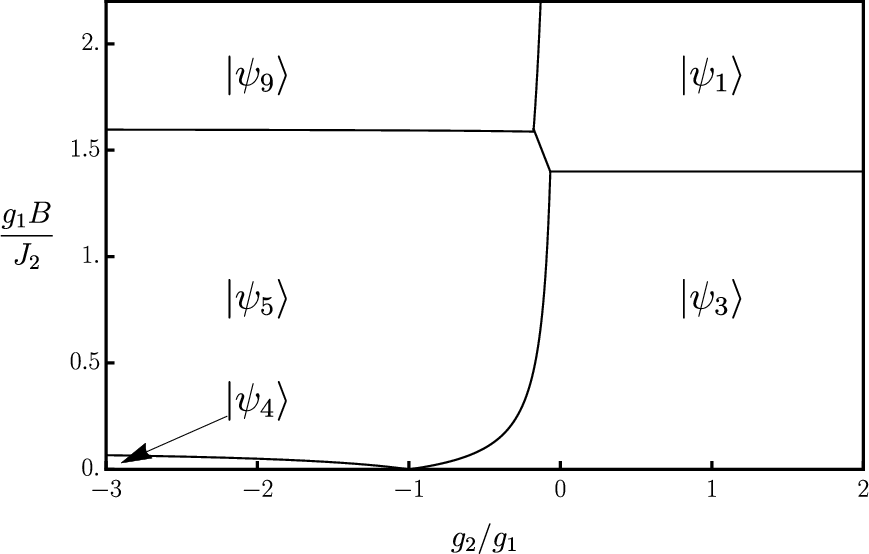} 
}
\vspace{-0.75cm}
\caption{ Ground-states phase diagram for $J_1=-1/5$, $J_2=1$}
\label{fig_9}
\end{figure}

\begin{figure}[h!]\label{fig9}
   \centering
  \resizebox{0.5\textwidth}{!}{
 \includegraphics{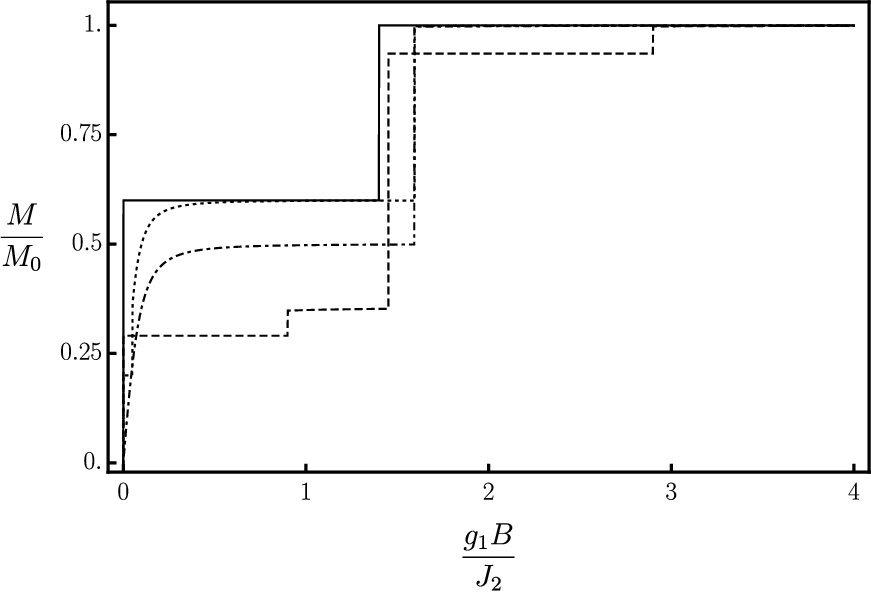} 
}
\vspace{-0.75cm}
\caption{ Magnetization processed for $J_1=-1/5$, $J_2=1$, $T/|J_1|=0.00001$, $g_1=1$ for different values of $g_2$: $g_2=2$ (solid); $g_2=-1/10$ (dashed); $g_2=-1$ (dot-dashed) and $g_2=-2$ (dotted).}
\label{fig_10}
\end{figure}

\begin{figure}
   \centering
  \resizebox{0.5\textwidth}{!}{
 \includegraphics{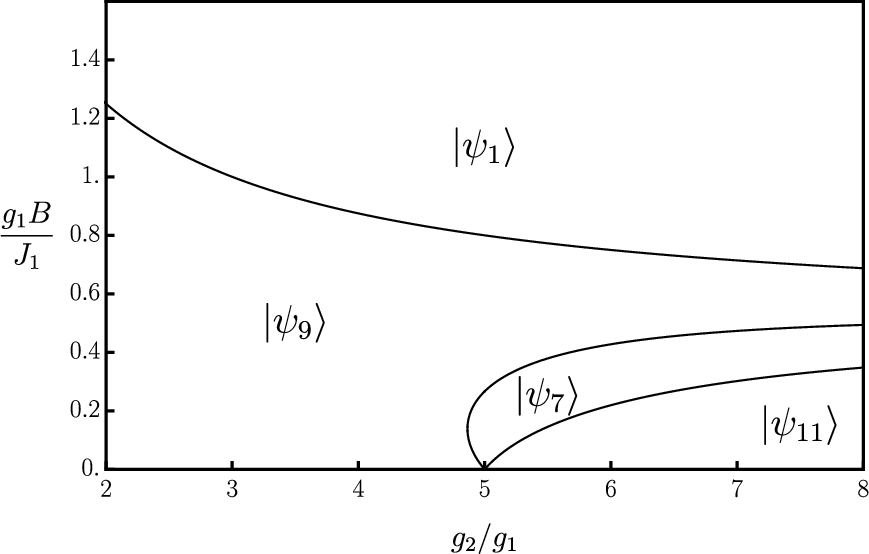} 
}
\vspace{-0.75cm}
\caption{ Ground-states phase diagram for $J_1=1$, and arbitrary $J_2 \leq 0$}
\label{fig_11}
\end{figure}

\subsubsection{Non-uniform antiferromagnetic couplings, $J_2>J_1>0$}

The case of antiferromagnetic non-uniform coupling with $J_2>J_1>0$ shares many features of the previous one (See Fig. \ref{fig_7} for $J_1=1/5$ and $J_2=1$). The set of possible ground states is the same, $|\psi_1\rangle$, $|\psi_3\rangle$, $|\psi_5\rangle$ and $|\psi_9\rangle$, again the zero-field ground state is non-degenerate and is given by $|\psi_5\rangle$. As in the previous case we plotted the ground states phase diagram for negative and positive values of $g_2$ with saturated state given by $|\psi_9\rangle$ and $|\psi_1\rangle$ respectively. However, the topology of diagram is a bit different as well as an additional feature appears here, for a narrow region of positive $g_2$ a direct transition from $|\psi_3\rangle$ to $|\psi_9\rangle$ is found. The magnetization processes, corresponding to different values of $g_2$ for this case are presented in Fig. \ref{fig_8}.

\subsubsection{Mixed case $J_{1} < 0$ and $J_{2} > 0$ }
The ground state phase diagram for the case $J_1=-1/5$ and $J_2=1$ is presented in Fig. \ref{fig_9}. Here we see additional possible ground state $|\psi_4\rangle$ along with all four ground states form the previous case. Interestingly, $|\psi_4\rangle$ is the zero-field ground state for negative $g_2\leq -1$. At the point $g_2=-1$ it is degenerate with the $g_2 \geq -1$ zero-field ground state, $|\psi_3\rangle$ and $|\psi_5\rangle$, which was the zero-field ground state for all range of $g_2$ in previous case. However, the region of $|\psi_4\rangle$ ground state in the phase diagram is extremely narrow. As usual, saturated state for positive values of $g_2$ is fully positively polarized $|\psi_1\rangle$ and for the negative values of $g_2$ the same role plays $S^z=1$ eigenstate $|\psi_9\rangle$. Typical magnetization curves for this case are presented in Fig. \ref{fig_10}.

\subsubsection{ Mixed case $J_{1} > 0$ and $J_{2} \leq 0$ }
\begin{figure}[h!]
   \centering
  \resizebox{0.5\textwidth}{!}{
 \includegraphics{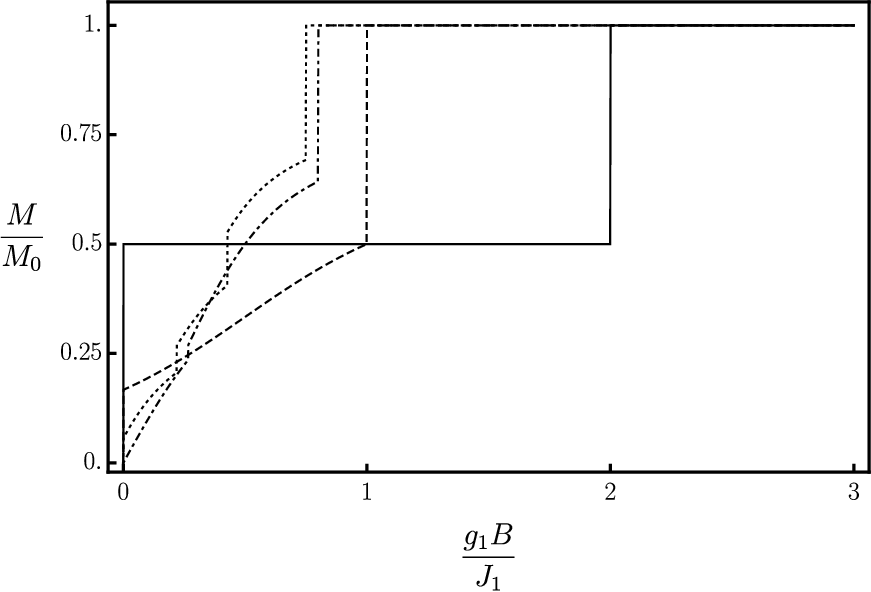} 
}
\vspace{-0.75cm}
\caption{\label{fig:phasediagram} Magnetization processed for $J_1=1$ and arbitrary $J_2\leq 0$, $T/J_1=0.00001$ and $g_1=1$ for different values of $g_2$: $g_2=1$ (solid); $g_2=3$ (dashed); $g_2=5$ (dot-dashed) and $g_2=6$ (dotted).}
\label{fig_12}
\end{figure}

Another case of mixed ferromagnetic-antiferromagnetic couplings corresponds to  $J_{1} > 0$ and $J_{2} \le 0$. The corresponding ground-states phase diagram is presented in Fig. \ref{fig_11}. Without loss of generality we set here $J_1=1$ and $J_2=-1$, however, the phase diagram preserve its structure for arbitrary $J_2 \le 0$. As was mentioned above, this is due to Eq. (\ref{J2_comm}). Here, the zero-field ground states can be $|\psi_9\rangle$, $|\psi_7\rangle$ and $|\psi_{11}\rangle$ depending on the value of $g_2$.  All the three ground states become degenerate at the particular value  of $g_2=5$. Saturated state is $|\psi_1\rangle$ for arbitrary value of $g_2$. The magnetization plots illustrating this phase diagram are presented in Fig. \ref{fig_12}.

\section{Negativity Plots}
In this section we present our main results concerning the manipulation of the quantum entanglement of the spins within the cluster by means of magnetic field. Particulary, the effects of non-conserving magnetization related to the non-uniform $g$-factors are in the focus of our attention.  There are several ways of quantifying the quantum entanglement using various entanglement measures \cite{ent, horodecki}. However, for the mixed-spin clusters negativity \cite{vidal} is the most convenient one, as it can be easily constructed and calculated for any pair of spins with the aid of reduced density matrix. For the system of three particles three pairwise entanglement measures can be considered. The negativity value, which varies from 0 (no entanglement) to $1/2$ (maximally entangled pair), corresponding to the i-th and j-th particles of the system,  $Ne_{ij}$, equals the sum of absolute values of negative eigenvalues of partially transposed reduced two-particle density matrix, $\rho_{ij}^T$, which is constructed in the following way:
\begin{eqnarray}
&&\left\langle \tilde{\xi_i}, \xi_j \right| \rho_{ij}^T \left| \xi_i, \tilde{\xi_j} \right\rangle= \left\langle \xi_i, \xi_j  \left| \rho_{ij} \right| \tilde{ \xi_i},
\tilde{\xi_j}\right\rangle, \\
&&\rho_{ij}=\sum_{\xi_k}\left\langle \xi_i, \xi_j, \xi_k \left| \rho \right|  \tilde{\xi_i}, \tilde{\xi_j}, \xi_k \right\rangle \nonumber
\end{eqnarray}
where, $\left|\xi_i, \xi_j, \xi_k \right\rangle$ is a standard basis for the states of $\mathbf{s_1}, \mathbf{s_2}$ ($\xi_i=\pm1/2$) and $\mathbf{S_3}$ ($\xi_i=-1, 0, 1$) spins. Then the negativity is obtained according to
\begin{equation}
Ne_{ij}=\sum\limits_{a}|\mu_{a}|,
\end{equation}		
where $\mu_{a}$ – negative eigenvalues of partially transposed reduced two-particle density matrix $\rho^{T}_{ij}$ \cite{vidal}. As the main object of our research is a quantum entanglement, i. e. the zero-temperature or the pure state properties , the density matrix we are working with is defined for each of the twelve eigenstates of the Hamiltonian as a pure-state density matrix:.
\begin{equation}
\rho_{i}=|\Psi_{i}\rangle \langle \Psi_{i}|, \;\; i=1,...,12.
\end{equation}
In case of $n$ degenerate eigenstates one should use
\begin{eqnarray}
\rho_{i_1...i_n}=\frac {1}{n}\sum_{a=1}^n|\Psi_{i_a}\rangle \langle \Psi_{i_a}| .
\end{eqnarray}
\begin{widetext}
\begin{figure}[ht]
\begin{center}
\includegraphics[width=18cm]{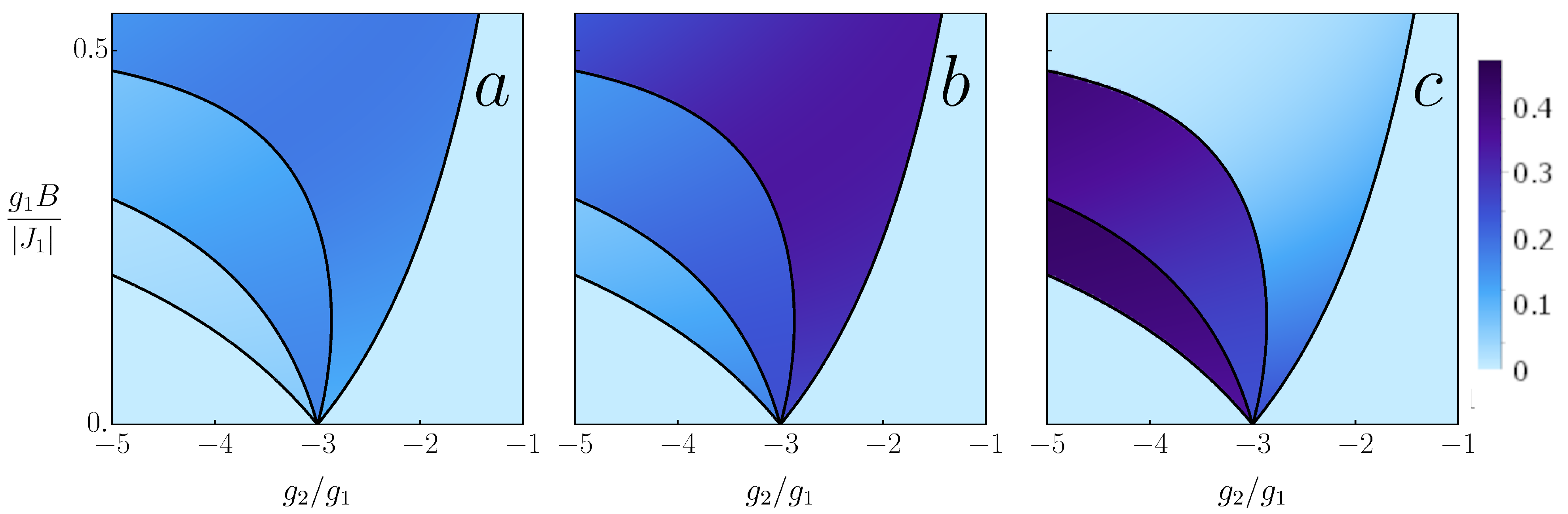}
 \end{center}
  \caption{Density plots of negativity for the case of $J_1<0, J_2\leq 0$. Panels (a), (b) and (c) show the values of $Ne_{12}$, $Ne_{23}$ and $Ne_{13}$ respectively.}
  \label{fig13}
\end{figure}
\end{widetext}
Therefore, in general thirty six negativity functions are relevant for the mixed-spin trimer, however, many of them are trivial, or just vanish. In addition to that, the case of $g_1=g_2$ should be considered separately. For the spin clusters with conserved magnetization eigenvectors do not contain magnetic filed dependent coefficient, thus neither do the negativity. However, all they can be obtained analytically, we list the expressions for non-zero negativity functions in the Appendix. Thus, for the eigenstates form $|\psi_5\rangle$ to $|\psi_{12}\rangle$ at least for some pairs of spins negativity may have continuous dependence on the magnetic field, in contrast to the constant values for the clusters with conserved magnetization. Interestingly, all negativity functions have no dependents on $J_2$. As was mentioned above, some eigenstates of the system with non-uniform $g$-factors do not admit continuous limit $g_2\rightarrow g_1$, for those eigenstates negativity was calculates separately. For demonstrating the overall picture of how quantum entanglement changes across the various eigenstates we present density plots of negativity projected into the ground-states phase diagrams (See Figs. \ref{fig13}-\ref{fig17}). In Figs. \ref{fig13}, \ref{fig14} and \ref{fig17} panel $(a)$ corresponds to $Ne_{12}$, $(b)$ to $Ne_{23}$ and $(c)$ to $Ne_{13}$. For two phase diagrams, Figs. \ref{fig15} and \ref{fig16} only two panels are presented, as for both cases, $J_2>0, J_1=-\frac 15 J_2$ and $J_2>0, J_1=\frac 15 J_2$, $Ne_{12}$ vanishes almost everywhere, except for small region within the high-field ground state $|\psi_9\rangle$ where ist value does not exceed $0.07$, for those plots panel $(a)$ corresponds to $Ne_{23}$ and panel $(b)$ to $Ne_{13}$. For the case of ferromagnetic couplings only negative $g_2$ can lead to non-trivial magneto-thermal and entanglement properties (See Fig. \ref{fig13}). Here one can see essentially entangled state of the first and the third spins (panel (c)) within the $|\psi_{11}\rangle$ eigenstate. To have more clear picture of how magnetic field affects the negativity, or how negativity is changing under the magnetization process we plotted the negativity plots as a function of magnetic field for $g_2=-4$ in Fig. \ref{fig18}. Three plots, corresponding to the entanglement of three pairs of spins, are composed of four parts, each one corresponding to a certain eigenstates form the sequence of transitions taking place during the magnetization process, $|\psi_2\rangle\rightarrow |\psi_{11}\rangle\rightarrow  |\psi_7\rangle\rightarrow |\psi_9\rangle$. Only negativity for $|\psi_2\rangle$ is zero for all pairs of spins, the other eigenstates feature explicit magnetic field dependence of negativity. Negativity reaches its maximum for the first and the third spins within $|\psi_{11}\rangle$ eigenstate and its value is approximately equal to $0.41$. Another situation when non-uniform $g$-factors bring to essential enhancement of entanglement is presented in Figs. \ref{fig14} and \ref{fig19}. The most efficiently non-conserving magnetization affects the entanglement of the system spins in case of uniform antiferromagnetic couplings, $J_1=J_2>0$. As was shown in  Fig. \ref{fig_5}, possible ground states of the model are $|\psi_1\rangle, |\psi_3\rangle, |\psi_5\rangle$ and $|\psi_9\rangle$. However, at uniform $g$-factors, $g_2=g_1$, $|\psi_3\rangle$ and $|\psi_9\rangle$ transform into $|\psi_3\rangle_0$ and $|\psi_9\rangle_0$ form Eq. (\ref{psi_0}). This fact is indicated in Fig. \ref{fig14} by a yellow line separating $|\psi_3\rangle$ and $|\psi_9\rangle$ at $g_2=g_1$. It is obviously seen from density plots of the negativity, that systems with $g_2\neq g_1$ exhibit much higher entanglement that their uniform counterpart, which is manifested in the most straightforward way for $Ne_{13}$ within the $|\psi_3\rangle$ eigenstate. Magnetic field behaviour of negativity for both uniform and non-uniform $g$-factors cases is presented in Fig. \ref{fig19}. As usual, panels (a), (b) and (c) corresponds to $Ne_{12}$, $Ne_{23}$ and $Ne_{13}$. For all panels solid and dot-dashed lines demonstrate the negativity variation under the magnetization process for $g_2=3/2$ and $g_2=-1/2$, respectively, whereas dashed line corresponds to the uniform $g$-factor case, $g_2=1$. For $Ne_{12}$ (panel (a)) no entanglement is observed for $g_2=3/2$, though during the magnetization process the system passes through the following sequence of eigenstates, $|\psi_5\rangle\rightarrow |\psi_3\rangle\rightarrow |\psi_1\rangle$, which have a complicated structure, but it turns out, that the first and the second spins are not entangled there.  At the same time uniform $g$-factors case (dashed) exhibits non-zero constant negativity for the finite range of magnetic field, corresponding to the phase boundary between $|\psi_3\rangle$ and $|\psi_9\rangle$, which in its turn is a degenerate superposition of $|\psi_3\rangle_0$ and $|\psi_9\rangle_0$. The negativity has a constant value here, $Ne_{12}^{(3+9)_0}=\frac 18 \left(\sqrt 2-1\right)\simeq 0.052$. For the case of $g_2=-1/2$ (dot-dashed) one can see a jump from zero to $Ne_{12}^9 \simeq 0.14$ with the further continuous decrease. Negativity plots quantifying entanglement for the second and the third particle can be seen in the panel (b). Interestingly, in the case of uniform $g$-factors (dashed) this pair of spins exhibit the strongest entanglement at low field, $Ne_{23}^5=1/3$, then it drops to another constant, but much smaller value, $Ne_{23}^{(3+9)_0}=\frac {1}{16}\left(\sqrt{17}-3\right)\simeq 0.07$. The case $g_2=3/2$ depicted by the solid line doesn't look really promising as only in the vicinity of $B=0$ the negativity is close to the uniform case, then it is decreasing a little bit, and after transition to $|\psi_3\rangle$ it drops down to zero. The case of negative $g_2=-1/2$ (dot-dashed) has no advantage over the uniform $g$ case for low field, actually within the $|\psi_5\rangle$ eigenstate, when $Ne_{23}^5$ drops down from $1/3$ to the value slightly bellow 0.2. However, it undergoes a jump at the transition point to the eigenstate $|\psi_9\rangle$ with the further slow decrease, being, nevertheless, much higher than the negativity of the uniform $g$ system.
\begin{widetext}

\begin{figure}[h]
\begin{center}
 \includegraphics[width=18cm]{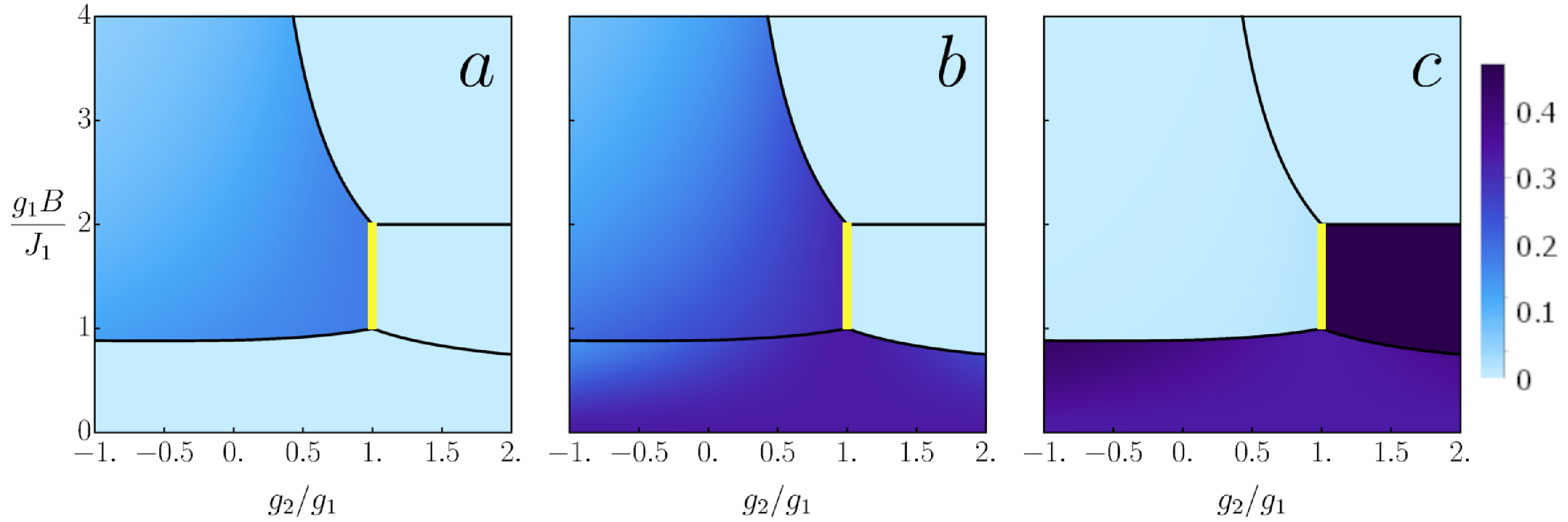}
  \caption{Density plots of negativity for the case of $J_1=J_2>0$. Panels (a), (b) and (c) show the values of $Ne_{12}$, $Ne_{23}$ and $Ne_{13}$ respectively.}
  \label{fig14}
  \end{center}
\end{figure}

\end{widetext}

The most significant enhancement of entanglement due to non-uniform $g$-factors occurs for the first and the third particles (panel (c)). The negativity curve for uniform $g$-factors has the same form as $Ne_{23}$, two plateaus at $1/3$ and $\frac {1}{16}\left(\sqrt{17}-3\right)\simeq 0.07$ corresponding to $|\psi_5\rangle$ and degenerate superposition of $|\psi_3\rangle_0$ and $|\psi_9\rangle_0$ respectively. Negativity curve for $g_2=3/2$ (solid) has weak magnetic field dependence within the $|\psi_5\rangle$ eigenstate where it demonstrates slow growth starting from $1/3$ value, but transition to the $|\psi_5\rangle$ eigenstate is accompanied with a jump to a constant value $Ne_{13}^5=\sqrt 3/2\simeq 0.47$ which is very close to maximal possible value, $1/2$. This is the highest value of negativity which can be achieved for the system under consideration. It is worth mentioning here, that the value of negativity, $Ne_{13}^3=\sqrt{3}/2$ is independent of $g$-factors values, however, for $g_2 \leq 1$ $|\psi_3\rangle$ does not appear as eigenstate during zero-temperature magnetization process. Thus, for the system with uniform antiferromagnetic exchange couplings, $J_1=J_2>0$, even an arbitrary small difference between $g_2$ and $g_1$ brings to essential increase of negativity $Ne_{13}$. The negativity curve for negative $g_2=-1/2$ (dot-dashed) demonstrates an enhancement of entanglement for $|\psi_5\rangle$ eigenstate with respect to uniform $g$ case, one can see monotone increase of $Ne_{13}^5$ up to approximately $0.41$, which corresponds to about 23$\%$ increase with respect to $Ne_{13}^5=1/3$ for $g_2=g_1$. For the system with mixed couplings, $J_1J_2<0$, as it can be seen from Figs. \ref{fig15}-\ref{fig17} properties of the negativity do not exhibit essential changes in case of non-uniform $g$-factors, or the eigenstates with enhanced entanglement can appear only for rather large value of the ration $g_2/g_1$ (Fig. \ref{fig17}, panel (c)).

\begin{figure}[h]
\begin{center}
 \includegraphics[width=\linewidth]{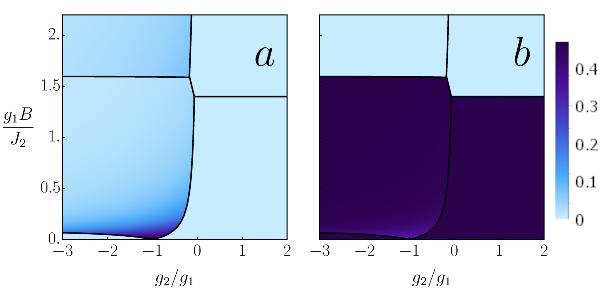}
  \caption{Density plots of negativity for the case of $J_2>0$ and $J_1=-\frac 15 J_2$. Panels (a) and (b) show the values of $Ne_{23}$ and $Ne_{13}$ respectively.}
    \label{fig15}
  \end{center}
\end{figure}

\begin{figure}[h]
 \begin{center}
 \includegraphics[width=\linewidth]{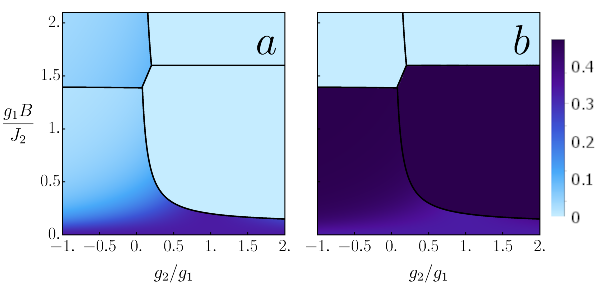}
  \caption{Density plots of negativity for the case of $J_2>0$ and $J_1=\frac 15 J_2$.  Panels (a) and (b) show the values of $Ne_{23}$ and $Ne_{13}$ respectively.}
  \label{fig16}
  \end{center}
\end{figure}

\begin{widetext}

\begin{figure}[h]
 \begin{center}
 \includegraphics[width=18cm]{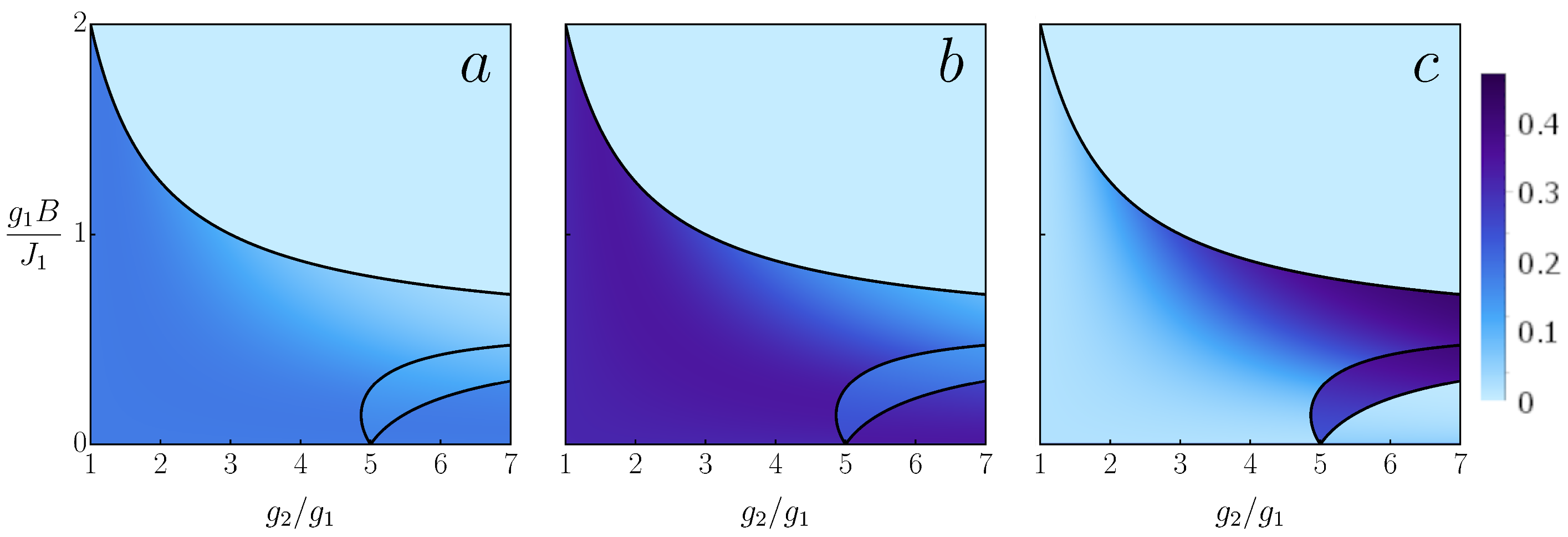}
  \caption{Density plots of negativity for the case of and $J_1>1$ and arbitrary $J_2<0$. Panels (a), (b) and (c) show the values of $Ne_{12}$, $Ne_{23}$ and $Ne_{13}$ respectively.}
  \label{fig17}
  \end{center}
\end{figure}
\end{widetext}

\begin{figure}[h!]
\includegraphics[clip=on,width=74mm]{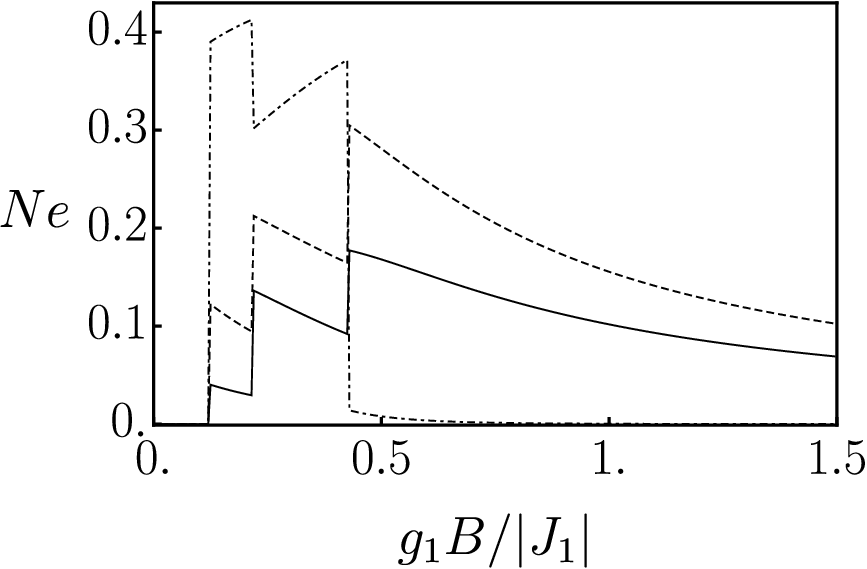}
\caption{Negativity magnetic field dependence for the case $J_1<0, J_2 \leq 0$, $g_1=1$ and $g_2=-4$. Solid line corresponds to $Ne_{12}$, dashed to $Ne_{23}$ and dot-dashed to $Ne_{13}$ }
\label{fig18}
\end{figure}

\begin{figure}
\centering
\includegraphics[clip=on,width=74mm]{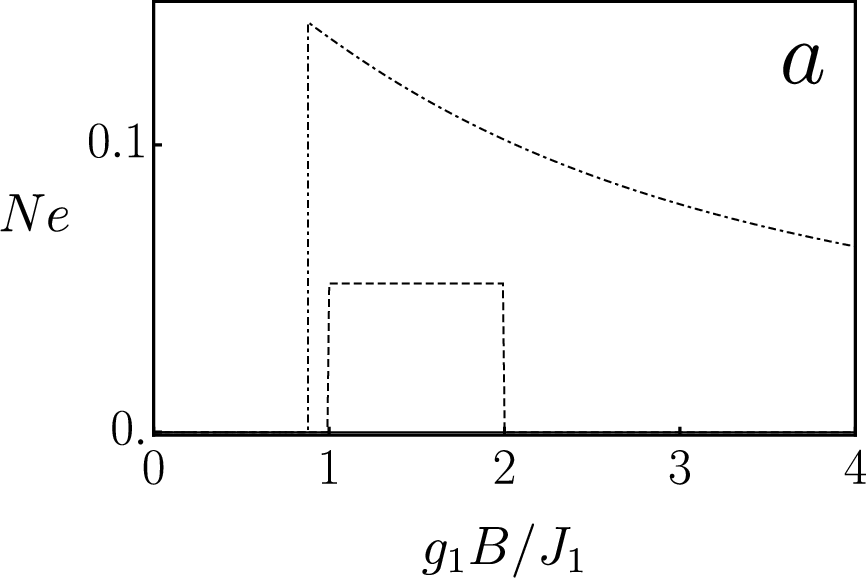}
\includegraphics[clip=on,width=74mm]{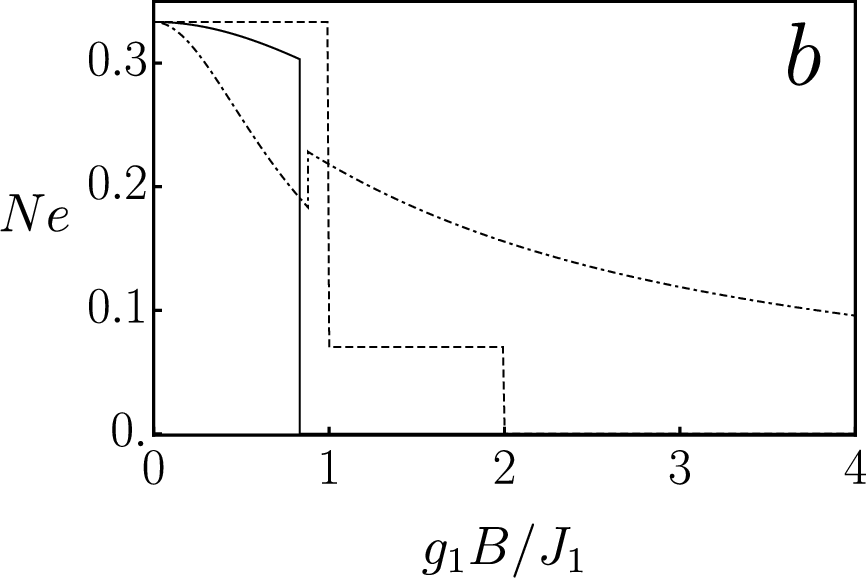}
\includegraphics[clip=on,width=74mm]{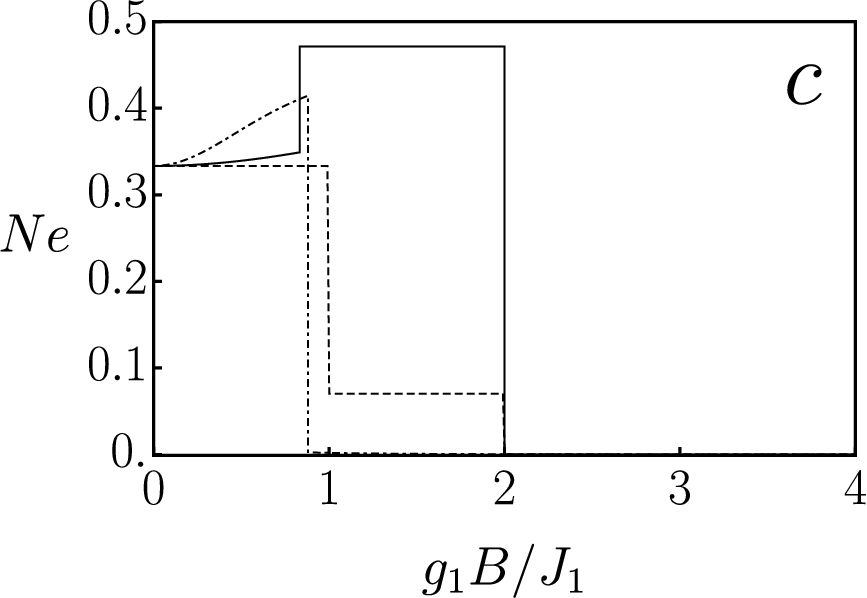}
\caption{Negativity magnetic field dependence for the case $J_1=J_2>0$. Panels (a), (b) and (c) show $Ne_{12}$, $Ne_{23}$ and $Ne_{13}$ respectively. In each panel solid line corresponds to $g_2=3/2$, dashed line corresponds to uniform-$g$ case, $g_2=g_1$, and dot-dashed line represent $g_2=-1/2$ case. }
\label{fig19}
\end{figure}

\section{Conclusion}
In the present paper we considered the problem of control and optimization of quantum entanglement in triangular-shaped mixed spin $(1/2,1/2,1)$ molecular magnet models by means of magnetic field when $g$-factors of spins are non-uniform. More specifically, we chose the model which provided us with the most simple analytical solutions due to high symmetry of the Hamiltonian. The parameters of the model turn out to distribute in the following way: spin-1 ion and one of the two spin-1/2 ion have the same $g$-factor, $g_1$, whereas another spin-1/2 ion has different $g$-factor, $g_2$; two spin-1/2 ion are coupled to spin-1 ion with different exchange constants, $J_1$ and $J_2$ and the coupling between two spin-1/2 ions is given by the same $J_1$ exchange constant. The presence of non-uniform $g$-factors makes the magnetization non-conserved which, in its turn, brings to a series of physical features, like asymptotic saturation, non-constant values of magnetization and negativity within the certain eigenstates. It also has deep impact on the entanglement properties, enhancement or weakening. We illustrated all these features by plotting ground states phase diagrams for five different ranges of exchange couplings and the corresponding magnetization plots. The main results, however, concern the entanglement properties and their modification under the presence of non-conserving magnetization. The most dramatic impact of non-conserving magnetization can be seen in the case of uniform antiferromagnetic couplings, $J_1=J_2 >0$, where one can have multiple increase of $Ne_{23}$.
This effect is qualitative and robust, for arbitrary small difference between $g_2$ and $g_1$ the value of $Ne_{23}^3\simeq 6.7 Ne_{23}^{(3+9)_0}\simeq 0.47$, which is just six percent less than the maximal possible value of negativity. Thus, molecular magnets with non-conserving magnetization offer an opportunity of more efficient manipulation of the quantum entanglement by means of magnetic field. The main difference from the uniform-$g$ case is a possibility to have continuous dependence of the negativity on the magnetic field within the same ground state as well as to achieve essentially strongly entangled eigenstates.  In our research we generated non-conserving magnetization by assuming non-uniformity of the $g$-factors, which can be understood in a direct way as presence of $4f$ or $5f$ ions along with $3d$ or/and $4d$ ions into the structure of trimetallic complex \cite{mm1,mm2,mm3,mm4}. However, there is another physical setting which can effectively bring to a problem of quantum spin clusters with the difference in local field acting on different spins in the cluster. They are the models in which small groups of quantum spins are assembled into the regular lattice by alternating with the groups of Ising spins \cite{sou20, str05, vis09, van10, bel14, oha09, bel10, oha12, bel13, zad23}. The later situation is realized in a series of coordination polymer compounds \cite{sou20, str05, vis09, van10, bel14}.

\section{Appendix}
Here we present the list of analytical expressions for negativities corresponding to three pairs of spins.
  \begin{eqnarray}\label{Negativity}
&& {Ne}_{12}^{3}={Ne}_{12}^{4}={Ne}_{23}^{3}={Ne}_{23}^{4}={Ne}_{12}^{5}=0,  \\
&& {Ne}_{13}^{3}={Ne}_{13}^{4}=\frac{\sqrt{2}}{3}, \nonumber\\
&&{Ne}_{23}^{5}=\frac{1}{6} \left|\frac{M_{+}^2-M_{+}\sqrt{M_{+}^4+8}}{M_{+}^2+1}\right| +\frac{1}{6} \left| \frac{1-\sqrt{8 M_{+}^4+1}}{M_{+}^2+1}\right|, \nonumber\\
&&  {Ne}_{13}^{5}=\frac{1}{6} \left|\frac{M_{+}^2-\sqrt{M_{+}^4+8}}{M_{+}^2+1}\right| +\frac{1}{6} \left| \frac{1-\sqrt{8 M_{+}^4+1}}{M_{+}^2+1}\right|, \nonumber \\
&&        {Ne}_{12}^{7}=\frac{1}{6} \left| \frac{K_-^2-\sqrt{K_-^4+56 K_-^2+16}+4}{K_-^2+4}\right|,        \nonumber \\
&&{Ne}_{23}^{7}=\frac{2}{3} \left| \frac{2-\sqrt{2} \sqrt{K_-^2+2}}{K_-^2+4}\right| +\frac{1}{3} \left| \frac{K_-^2-\sqrt{K_-^4+8 K_-^2}}{K_-^2+4}\right|, \nonumber \\
&&  {Ne}_{13}^{7}=\frac{1}{3}\left| \frac{4-\sqrt{2} \sqrt{K_-^4+8}}{K_-^2+4}\right| +\frac{1}{3} \left| \frac{K_-^2-\sqrt{K_-^4+32}}{K_-^2+4}\right|,              \nonumber \\
&&         {Ne}_{12}^{9}=\left| \frac{1-\sqrt{G_+^2+1}}{G_+^2+3}\right|,
       \nonumber \\
&&      {Ne}_{23}^{9}=\frac{1}{2} \left| \frac{1-\sqrt{8 G_+^2+1}}{G_+^2+3}\right|,
           \nonumber \\
&&      {Ne}_{13}^{9}=\frac{1}{2} \left| \frac{G_+^2-\sqrt{G_+^4+8}}{G_+^2+3}\right|,
           \nonumber \\
&& {Ne}_{12}^{11}=\left| \frac{1-\sqrt{U_-^2+1}}{U_-^2+3}\right|,
              \nonumber \\
\end{eqnarray}
\begin{eqnarray}\nonumber
&&             {Ne}_{23}^{11}=\frac{1}{2} \left| \frac{1-\sqrt{8 U_-^2+1}}{U_-^2+3}\right|,
    \nonumber  \\
&&            {Ne}_{13}^{11}=\frac{1}{2} \left| \frac{U_-^2-\sqrt{U_-^4+8}}{U_-^2+3}\right|,
     \nonumber \\
&& {Ne}_{12}^{(3+9)_{0}}=\frac 18 \left(\sqrt 2-1\right),
               \nonumber  \\
&& {Ne}_{23}^{(3+9)_{0}}={Ne}_{13}^{(3+9)_{0}} =\frac{1}{16}\left(\sqrt{17}-3\right).
              \nonumber
\end{eqnarray}
Expressions for $Ne_{ij}^6$, $Ne_{ij}^8$, $Ne_{ij}^{10}$ and $Ne_{ij}^{12}$ have the same structure as $Ne_{ij}^5$, $Ne_{ij}^7$, $Ne_{ij}^{9}$ and $Ne_{ij}^{11}$ where $M^+, K^-, G^+$ and $U^-$ are replaced by $M^-, K^+, G^-$ and $U^+$ respectively.
\section{Acknowledgements}
 The authors express their gratitude to Vahag Abgaryan for useful discussions. They also acknowledge partial financial support form ANSEF (Grants No. PS-condmatth-2462 and PS-condmatth-2884) and from CS RA MESCS (Grants No. 21AG-1C047, 21AG-1C006, 20TTAT-QTc004 and 23AA-1C032).


\end{document}